\documentclass[]{article}
\usepackage{amsmath}
\usepackage{amsthm}
\usepackage{amssymb}
\usepackage{enumitem}
\usepackage{url}
\newtheorem{theorem}{Theorem}
\theoremstyle{plain}
\newtheorem{lemma}{Lemma}[section]
\newtheorem{proposition}[lemma]{Proposition}

\newtheorem{claim}[lemma]{Claim} 
\newtheorem{corollary}[lemma]{Corollary}

\theoremstyle{definition}

\newtheorem{Definition}[lemma]{Definition}
\newcommand{\customqed}[1]{{\renewcommand{\qedsymbol}{#1}\qed}}
\newcommand{\varqed}{\customqed{\hbox{$\lrcorner$}}}
\newenvironment{definition}{\begin{Definition}}{\varqed\end{Definition}}

\newtheorem{Remark}[lemma]{Remark}

\title{Kolmogorov-Loveland betting strategies lose the Betting game on open sets}
\author{Tomislav {Petrovi\'c}}

\begin{document}
\bibliographystyle{plain}
\maketitle

\begin{abstract}
Whether Kolmogorov-Loveland randomness (KLR) is the same as Martin-L\"of randomness (MLR) is a major open problem in the study of algorithmic randomness.
More general classes of betting strategies than Kolmogorov-Loveland ones have been studied in \cite{MMS, Rute, TP}. In each case it was proven that the class
induces a notion of randomness equivalent to MLR.
In all of those proofs, it was shown that the class
contains a finite set of betting strategies such that
for any given bound, when betting on a binary sequence contained in an effective open set
of small enough measure, at least one of the betting strategies in the set 
earns capital larger than the bound.
We show that the class of Kolmogorov-Loveland betting strategies does not have this property.
\end{abstract}

\section{Introduction}
Kolmogorov-Loveland randomness was first defined in \cite{Muchnik} in terms of betting strategies that bet on the values of bits of an infinite binary sequence.
Based on the values of bits it has seen so far, a betting strategy makes a bet by choosing a position to bet on next and placing some fraction of its capital on a guessed value of the bit at the chosen position as a wager. The value of the bit is then revealed, and the capital is updated. If the guess was wrong the wager is lost. If the guess was correct, the capital is increased by the wagered amount.
A \textit{monotonic betting strategy} must choose the positions on which to bet in increasing order, while the \textit{non-monotonic betting strategies} can choose the positions in any order, adaptively (depending on the bits that were revealed so far).
A betting strategy \textit{wins} on the binary sequence if the supremum of capital over the sequence of bets is unbounded.
A sequence is Kolmogorov-Loveland random (KLR) if no computable non-monotonic betting strategy wins on the sequence.

Computable non-monotonic betting strategies are also called \textit{Kolmogorov-Loveland betting strategies}. 
Both Kolmogorov in \cite{KKL} and Loveland in \cite{LKL}
considered non-monotonic choice of positions.
However, they
looked at the \textit{frequency} of $1$s over the chosen positions to define a
version of randomness
that is nowadays called \textit{Kolmogorov-Loveland stochasticity}.

In this paper, we will consider only Martin-L\"of randomness (\cite{ML}) with respect to the uniform (Lebesgue) measure, $\lambda$.
With this in mind, we give a brief definition.
An effective open set is a set of sequences that have a prefix in a computable enumeration of strings.
A Martin-L\"of randomness test is a computable enumeration of effective open sets called levels of the Martin-L\"of test. The $k$th level of the test has measure less than $2^{-k}$, and a sequence fails the test if it is contained in every level of the test.
A sequence is Martin-L\"of random (MLR) if it passes every Martin-L\"of randomness test.


Whether the set of KLR sequences is the same as the set of MLR sequences is a well studied open question in the field of algorithmic randomness \cite{LiV19}.

A set of betting strategies is called \textit{universal} if for every non-MLR sequence it contains a strategy that wins on the sequence. 
We can now restate "is KLR=MLR?" as 
"is the set of Kolmogorov-Loveland betting strategies universal?".
In fact, by a result in \cite{MillerMerkle}, we can consider just the total computable non-monotonic betting strategies since every partial strategy can be replaced by two total ones that win on the same sequences as the partial one.


For any betting strategy, making a bet consists of partitioning a set of sequences into two subsets and placing a wager on one of them.
The outcome of a bet determines which subset contains the sequence the strategy is betting on. 
A non-monotonic strategy,  when making a bet, partitions a set of sequences  into two sets according to the value of the bit at a chosen position.
In \cite{MMS}, more general betting strategies than the non-monotonic ones have been studied. 
A \textit{general betting strategy} has more freedom in partitioning sets.
When making a bet, the set determined by the outcome of a previous bet, $v$, can be split into any two chosen clopen sets $v_0,v_1$. The outcome of the bet determines in which of the two sets the sequence is, and the capital is updated in the following way. 
Suppose that the betting strategy wagers amount $w$ of current capital $c$ on $v_0$.
If the sequence the strategy is betting on is in $v_1$,
the updated capital is $c_1= c-w$. If the sequence is in $v_0$, the updated capital is $c_0=c+w\frac{\lambda(v_1)}{\lambda(v_0)}$.
Note that the measure of $v_0,v_1$ does not have to be equal.
It was shown in \cite{MMS} that there is a single total computable general betting strategy that is universal.

The outcomes of an infinite sequence of successive bets determine 
an infinite sequence of nested sets. 
If the measure over the sequence of sets for some infinite succession of bets does not go to zero, we say that a betting strategy has \textit{atoms}.
It is easy to see that a universal betting strategy necessarily has atoms.
An atomless computable betting strategy cannot be universal, since the sequence of outcomes of bets where wager was lost each time (to wit, the losing streak) then defines a ML-test - the sequences that fail the test at every level are non-MLR by definition, but the betting strategy does not win on them.

However, we can use van Lambalgen's theorem \cite{vLamb}
to show that there are two atomless computable general betting strategies $A$ and $B$ that are universal (by Alexander Shen, personal communication).
Let us encode a pair of sequences $(\alpha,\beta)$ as a single sequence $\gamma$ with $\alpha$ at even positions and $\beta$ at odd positions.
The betting strategy $B$ bets on odd positions like the universal betting strategy and, to be atomless, ocasionaly reads (wagers 0 capital on) even positions.
The betting strategy $A$ bets on the even positions like the universal betting strategy with an oracle $\beta$, with the bits of the oracle obtained by reading from the odd positions. By van Lambalgen's theorem, $\gamma$ is non-MLR if $\beta$ is non-MLR
(in which case strategy $B$ wins on $\gamma$), or, if $\alpha$ is non-MLR with $\beta$ as an oracle (in which case strategy $A$ wins on $\gamma$).

Furthermore, if we look more closely at the proof of Theorem 5.9. in \cite{vLamb},
there is a pair of computable betting strategies $A$ and $B$ such that,
given $n$,
for every sequence in the $n$th level of the universal Martin-L\"of test, 
$O_n$ (or any other given effective open set of measure less than $2^{-n}$),
either strategy $A$ has supremum of capital larger than $(n+1)^2$ when betting on the sequence or strategy $B$ has supremum of capital larger than $2^n/(n+1)^2$.

Thus, the class of atomless betting strategies (called exhaustive betting strategies in \cite{Rute}) contains a pair that, for a given bound on capital,
when betting on a binary sequence in an effective open set of small enough measure,
at least one of the betting strategies earns capital larger than the bound.
Similarly, one can show the same for the class of balanced (\cite{Rute}) or half-betting strategies (\cite{TP}). The class of general betting strategies, 
and, similarly, martingale processes of \cite{HitchcockLutz}, \cite{MMS}, contains a single betting strategy with this property.

We contrast this with the main result of this paper.
For every Kolmogorov-Loveland betting strategy 
we can compute a bound on the capital and 
construct an effective open set
of arbitrarily small measure that contains a sequence
such that 
every Kolmogorov-Loveland betting strategy, while betting on the sequence, 
will always have capital below its bound.

\section{Notation}
\newcommand{\sqs}{\widetilde}
\newcommand{\lplus}{{\lambda^+}}
\newcommand{\bbN}{\mathbb{N}}
The set of (finite binary) strings is denoted with $\{0,1\}^*$, the set of strings of length $\ell$ with 
$ \{0,1\}^\ell$, and the set of (infinite binary) sequences with 
$\{0,1\}^\infty$.
The length of string $s$ is denoted with $|s|$. The empty string, $\epsilon$, has length $0$. That is, $|\epsilon|=0$.
If a string $s$ is a prefix of string (or sequence) $s'$
we write $s\preceq s'$.
The binary value of a sequence $\sigma$ at position $p$ is denoted with $\sigma_p$.
A set of sequences prefixed by some string $s$ is called a basic set and is denoted with
$\sqs s$.
A union of basic sets is called an open set, its complement a closed set,
and a union of finitely many basic sets is a clopen set (both open and closed).
A union of disjoint sets is denoted with symbol $\sqcup$ .

A restriction $r$ is a sequence of symbols $0$, $1$ or $*$.
The set of restrictions is denoted with $\{0,1,*\}^\infty$.
We denote the symbol at position $p$ in restriction $r$ with $r_p$.
An infinite binary sequence $\sigma$ is consistent with a restriction $r$
if at every position $p$ where $r_p\neq *$, $\sigma_p=r_p$.
We denote the set of sequences consistent with restriction $r$ with
$\sqs r$, it is a closed set. For a set of restrictions $R$, $\sqs R$ denotes the set of sequences that are consistent with any restriction in $R$.

If the restriction $r$ has $*$ at all but finitely many positions,
we say that $r$ is a finite restriction.
Note that in this case $\sqs r$ is clopen.
The empty restriction, $\zeta$, has $*$ at all positions, 
and is consistent with every sequence.

\begin{definition}
	We denote with $\lambda$ uniform (Lebesgue) measure on $\{0,1\}^\infty$.
	Furthermore, we will be measuring only basic sets and their unions.
	Let $s$ be a string, the measure of the basic set $\sqs s$ is $2^{-|s|}$.
	Let $X$ be a union of disjoint basic sets, the measure of $X$ is the sum of measures of the basic sets.
\end{definition}
\begin{definition}
	Let $\rho$ be a partial function from strings to finite restrictions with these properties:
	\begin{itemize}
		\item The empty string, $\epsilon$, is mapped to the empty restriction, $\zeta$.
		\item If for some string $s$ and bit $b$, $\rho(sb)$ is defined then both $\rho(s)$ and $\rho(s\bar{b})$ are defined.
		These three restrictions are the same except at one position $p$ where $\rho(s)_p=*$, $\rho(s0)_p=0$ and
		$\rho(s1)_p=1$.
	\end{itemize}
	We will call $\rho$ a \textit{restriction function}.
	
	Let $\mu$ be a partial function from strings to non-negative reals such that:
	\begin{itemize}
		\item The empty string, $\epsilon$, is mapped to 1.
		\item If for some string $s$ and a bit $b$, $\mu(sb)$ is defined then both $\mu(s)$ and $\mu(s\bar{b})$ are defined, and $\mu(s)=\mu(sb)+\mu(s\bar{b})$.
	\end{itemize}
	We will call $\mu$ a \textit{mass function}.
	If $\mu$ is defined on all strings on which $\rho$ is defined, we call the pair $(\rho,\mu)$ a \textit{non-monotonic betting strategy}.
	We will say that the strategy is finite if the restriction function is defined for only finitely many strings.
	The strategy is computable if both $\rho$ and $\mu$ are computable.
	The \textit{Kolmogorov-Loveland betting strategies} are precisely the computable non-monotonic betting strategies.
	
	For a string $s$, and a betting strategy $(\rho,\mu)$ we call the pair 
	$(\rho(s),\mu(s))$ \textit{the betting outcome} for $s$. 
	A pair of betting outcomes for $s0,s1$ 
	is called \textit{a bet for} $s$.
	
	Sometimes it is more convenient to look at the ratio between mass assigned to $s$ and the measure of the set of sequences consistent with the restriction assigned to $s$, called the \textit{capital} for $s$ and denoted with $c$. That is,
	$c(s)=\mu(s)/\lambda(\tilde{\rho}(s))=2^{|s|}\mu(s)$.
	
	The maximum of capital over all prefixes of $s$ is the
	\textit{maximum of capital} for $s$ and we denote it with $\bar{c}$.
	That is,
	$$\bar{c}(s)=\max_{s'\preceq s} c(s')$$. 
	
	The \textit{maximal capital} that the strategy achieves when betting on $\gamma$, denoted $\hat{c}(\gamma)$, is the supremum of capital over the bets the strategy makes when betting on $\gamma$, that is, 
	$$\hat{c}(\gamma)=\sup_{\{s:\gamma\in\sqs\rho(s)\}} \bar{c}(s)$$.
	We say that the strategy \textit{wins} on $\gamma$ if $\hat{c}(\gamma)$ is unbounded.
\end{definition}

We can define a betting strategy incrementaly, bet by bet.

\begin{definition}
	Let $B$ be a non-monotonic betting strategy.
	Let $s$ be a string for which the outcome is defined and the outcomes for $s0,s1$ are not defined.
	We will call $s$ a \textit{leaf-string} of $B$ and
	the restriction $\rho(s)$ a \textit{leaf-restriction}.
	Let $B'$ be a non-monotonic betting strategy that is the same as $B$ for all strings, except that, additionally, it has defined outcomes for strings $s0,s1$.
	We will say that $B'$ is obtained from $B$ \textit{by defining a bet} for the leaf-string $s$.
	Let $B'$ be a strategy obtained from $B$ by 
	successively defining finitely many bets, we will denote this with
	$B\to B'$.
	
	We call a betting strategy that has a defined outcome only for the empty string
	an \textit{initial betting strategy}.
\end{definition}

Note that any betting strategy can be defined by starting with an initial betting strategy, and then successively defining bets.

We next define effective open sets.

\begin{definition}
	An open set $O\subseteq\{0,1\}^\infty$ is \textit{effective} if
	$O=\bigcup_i \sqs f(i)$,
	where $f$ is a computable map from natural numbers to strings.
\end{definition}

\section{Preliminaries}
We can now state the main result of this paper.
\begin{theorem}[Main Theorem]\label{main_result}
	For each Kolmogorov-Loveland betting strategy there is a computable bound and
	for all $k$ there is an effective open set of measure less than $2^{-k}$
	such that
	the set contains a sequence on which the maximal achieved capital of any
	Kolmogorov-Loveland betting strategy is below the strategy's bound.
\end{theorem}
%
%

We will prove the main result by introducing a game 
called the \textit{Betting game on open sets} that is played
between two players, the \textit{Chooser} and the \textit{Gambler}. 
We show that if there is a computable winning strategy in the game for the Chooser then this implies the main result. 

\newcommand{\bB}{\textbf{B}}
\begin{definition}
	The \textit{Betting game on open sets} is played between two players called \textit{the Chooser} and \textit{the Gambler} and has two parameters,
	a set of bounds on the maximal capital $H=\{h_1,h_2,
	\dots\}$ 
	and a size parameter $k$.
	
	The Chooser will be choosing clopen sets, and in the entire game, the Chooser must choose a sequence of clopen sets such such that their union (an open set) has measure less than $2^{-k}$.
	
	The Gambler starts the game with a countable set of initial non-monotonic betting strategies,
	denoted $\bB_0=\{B_1^0, B_2^0,\dots\}$.
	As the game progresses, the Gambler will be defining bets for the betting strategies
	with the goal to define strategies so that,
	for every sequence in the chosen clopen sets,
	there is at least one $n$ such that the $n$th betting strategy
	achieves a maximal capital of strictly more than the bound $h_n$
	when betting on the sequence.
	
    The game is played in turns.
    Let $B_n^{i-1}$ denote the Gambler's $n$th betting strategy at the beginning of $i$th turn,
    and let $\bB_{i-1}$ denote the entire set of Gambler's betting strategies at the beginning of $i$th turn, that is, $\bB_{i-1}=\{B^{i-1}_1,B^{i-1}_2,\dots\}$.
    We write $\bB_{i-1}\to \bB_i$ to denote that
    for all $n$, we have  $B_n^{i-1}\to B_n^i$.

	In $i$th turn of the game, the Chooser chooses a clopen set $C_i$ (can be empty) and reveals it to the Gambler. The Gambler then decides if he will define a new bet for one of his strategies and if so reveals this strategy and the new bet to the Chooser. Alternatively, the Gambler decides to define no new bets, and reveals this to the Chooser (in this case $\bB_{i-1}=\bB_i$). 
	
	We will say that the Gambler achieves the $i$th goal if in some turn (say $j$th), for every sequence in the chosen set $C_i$ there is some $n$
	such that the betting strategy $B^j_n$ achieves maximal capital larger than $h_n$
	when betting on the sequence.
	
	The Gambler \textit{wins} the game if every goal is eventually achieved, otherwise the Chooser wins.
	
	The Chooser has a \textit{winning strategy} for the Betting game, if there is some $H$ such that for any Gambler and every  parameter $k$
	the Chooser wins the game.
	We will say that the Chooser has a 
	computable winning strategy for the Betting game if it has a winning strategy
	and for every $n$ the capital bound $h_n$ is computable and 
	for all $i$ the
	clopen set of sequences $C_i$ is computable from $\bB_{i-1}$.
\end{definition}

\begin{lemma}\label{lem_cws4Basic=>main_res}
	If the Chooser has a computable winning strategy in the Betting game, this implies the main result, Theorem \ref{main_result}.
\end{lemma}
\begin{proof}
	Let $K=\{B_1,B_2,\dots\}$ be the set of Kolmogorov-Loveland betting strategies.
	From $K$, we will define a Gambler $G$.
	The Gambler $G$ starts the game with initial betting strategies
	$\{B'_1,B'_2,\dots\}$.
	In each turn,
	the Gambler $G$ runs the dovetailed computation of strategies in $K$ for one step and if in that step
	a betting strategy (say $B_n$) defines a bet then $B'_n$ defines that same bet,
	and if not then $G'$ does not define any new bets in this turn.
	
	To prove the lemma we show that
	\begin{claim}\label{compactness}
		If the Gambler $G'$ loses the Betting game with
		parameters $H=\{h_1,h_2,\dots\}$ and $k$ against a Chooser, 
		then there is
		a sequence in some chosen basic set such that for all $n$
		the maximal achieved capital of $B_n\in K$ is less than $h_n$. 
	\end{claim}
	\begin{proof}[Proof by compactness.]
		Let $(\rho_n,\mu_n)$ denote the restriction and mass function of the $n$th strategy, $B_n$, 
		and let $c_n$, $\bar{c}_n$, $\hat{c}_n$ denote its capital, maximum capital and maximal achieved capital functions, respectively. 
		
		If for some sequence $\gamma$, $\hat{c}_n(\gamma)>h_n$ then there is a shortest string $s$ such that $\gamma$ is consistent with $\rho_n(s)$ and $\bar{c}_n(s)$ ($=c_n(s)$) is strictly larger than the capital bound for the $n$th strategy, $h_n$.
		Let $s_1,s_2,\dots$ be the effective enumeration of such strings and let $O_n$ be the effective open set that contains sequences from clopen sets $\sqs\rho_n(s_1),\sqs\rho_n(s_2),\dots$.
		Let $O$ be the (effective) open set that is the union of $O_1,O_2,\dots$.
		
		By compactness, for any clopen set $C$ that is a subset of $O$ there
		are finitely many basic sets in the enumeration of $O$ such that their union contains $C$. 
		In particular this is also true for the chosen clopen sets.
		Let $C_i$ be the clopen set chosen in the $i$th turn. 
		If for every sequence in $C_i$,
		some strategy from $K$ achieves maximum capital higher than its bound, then this happens after finitely many betting strategies have
		made finitely many bets, and the Gambler $G$ will in some turn define all of those bets for all of those strategies and achieve the $i$th goal.
		
		Therefore, the Gambler $G$ loses only if 
		there is some turn $i$ and a sequence in $C_i$ 
		such that 
		for all $n$ the maximal achieved capital of 
		the $n$th betting strategy in $K$  is less than $h_n$.
	\end{proof}
	 
	 If the Chooser has a computable winning strategy, there is some computable set of capital bounds $H$ such that for any $k$,
	 the set of chosen sequences
	 is an effective open set 
	 of size less than $2^{-k}$. 
	 This set, by Claim \ref{compactness}, 
	 contains a sequence for which the
	 maximal achieved capital of the $n$th strategy in $K$ is less than $n$th bound in $H$, for all $n$.
	 
\end{proof}

\begin{definition}
	Let a non-monotonic betting strategy $(\rho,\mu) $ have the following property:
	for all strings $s$, if the betting outcome is defined for $s$ then the difference between maximal capital for $s$ and capital for $s$ is less than $2$, that is, $c(s)\geq \bar{c}(s)-2$.
	We say that such a betting strategy is \textit{conservative}.
	
	We will say that the Gambler in the Betting game is \textit{conservative} if he defines bets so that the betting strategies defined by those bets are conservative.
\end{definition}

For every betting strategy $B$, we can construct a conservative betting strategy $B'$ that has the same restriction function, and has a mass function such that the maximum of capital of $B'$ is logarithmic in the maximum of capital of $B$. The construction is in \cite{BuhrmanMelkebeek} attributed to \cite{Mayordomo_phd}, and in other papers is often referred to as winning "slowly-but-surely", and the strategy $B'$ as $B$ "with savings".
For completeness we will also give the definition here. 

\begin{definition}\label{conservative_ver}
Let $B=(\rho,\mu)$ be a betting strategy and let $c$ denote its capital function,
that is, for any $s$ for which the outcome is defined we have $\mu(s)=c(s)\lambda(\tilde{\rho}(s))$.

From $B$ we will construct a strategy $B'$,
that has a different capital function, $c'$.
The capital $c'$ is defined as a sum of two functions, $c_B$ (the capital "in the bank") and $c_P$ (the capital "for play").
For the empty string the capital "for play" is $1$, and the capital "in the bank" is $0$.
That is, $c_P(\epsilon)=1$, $c_B(\epsilon)=0$.
The capital in the bank is never used for betting, the wagered capital comes out of the capital "for play".
The fraction of the capital "for play" that is wagered in a bet is the same as the fraction of the entire capital, $c$, of the original strategy that gets wagered in a bet,
and then, depending on the outcome, the "for play" capital gets increased or decreased.
As soon as the "for play" capital becomes larger than $2$, half of it is transfered to "the bank".

More formally, for any $s$ for which a bet is defined in the original betting strategy $B$,
let $f$ denote the fraction of capital that is wagered, $w$ the wagered amount, and $b$ the guessed value of the bit, $f = w/c(s)$, $w=c(sb)-c(s)$.
Then 
$$c_P(sb)=\begin{cases}
	\frac{(1+f)}{2} c_P(s) & \text{ if } (1+f) c_P(s) \geq 2\\
	(1+f) c_P(s)   & \text{ otherwise }
\end{cases}$$
$$
c_B(sb)
\begin{cases}
	c_B(s)+\frac{(1+f)}{2} c_P(s) & \text{ if } (1+f) c_P(s) \geq 2\\
	c_B(s)   & \text{ otherwise }
\end{cases}
$$
and
$$c_P(s\bar{b})= (1-f) c_P(s)$$ 
$$c_B(s\bar{b})=c_B(s)$$.

Let $c'(s)=c_P(s)+c_B(s)$ and $\mu'(s)=c'(s)\lambda(\tilde{\rho}(s))$.
We will say that the betting strategy $B'=(\rho,\mu')$ 
is the betting strategy $B=(\rho,\mu)$ \textit{with savings}.

For any $s$ for which a bet is defined in the original strategy $B$,
we will say that the pair of betting outcomes
$(\rho(s0),\mu'(s0)),(\rho(s1),\mu'(s1))$ is a conservative version of the original bet
$(\rho(s0),\mu(s0)),(\rho(s1),\mu(s1))$.
\end{definition}

\begin{lemma}\label{slowly-but-surely}
	Let $B$ be a betting strategy and let $B'$ be $B$ with savings.
	The strategy $B'$ is conservative, and
	for all strings $s$ for which the outcome is defined in $B$,
	we will have that 
	$c'(s)> \log \bar{c}(s)-2$.
\end{lemma}
\begin{proof}
	Let the capital of strategy $B'$ be $c'(s) = c_P(s)+c_B(s)$
	as in Definition \ref{conservative_ver}.
	We have that $c'(s)\geq c_B(s)$. 
	From the Definition \ref{conservative_ver}, we have that for any $s$,
	$c_P(s)\in [0,2)$.  
	Furthermore, since $c_B$ is non-decreasing,
	we have that for the maximum capital for $s$,
	$\bar{c}'(s)< c_B(s)+2$.
	Therefore
	$\bar{c}'(s)< c'(s)+2$. That is,
	the strategy $B'$ is conservative.
	
	For the original strategy, $B$, and any sequence $\sigma$,
	let $s_k$ denote the shortest prefix of $\sigma$ such that
	$c(s_k)\geq2^k$ (if such prefix exists). Since $s_k$ is the shortest prefix for which the capital is larger than the bound, $\bar{c}(s_k)=c(s_k)$.
	
	For $s\prec s_1$ we will have that
	$c_P(s)=c(s)$ and $c_B(s)=0$.
	Assume that for some $k$ 
	we have $c_P(s_k)=c(s_k)2^{-k}$ and $c_B(s_k)=\sum_{i\leq k} c_P(s_i)$.
	For all $s_k\preceq s\prec s_{k+1}$
	we will have 
	$c_P(s)= c_P(s_k)\frac{c(s)}{c(s_k)}=
	c(s)2^{-k}$ and $c_B(s)=c_B(s_k)$.
	For $s_{k+1}$, the capital $c(s_{k+1})\geq 2^{k+1}$, the "for play" capital becomes larger than $2$, and half of it is transferred to "the bank", that is $c_P(s_{k+1})=\frac{1}{2}c(s_{k+1})2^{-k}$ and 
	$c_B(s_k)=c_B(s_k)+ c_P(s_{k+1})$.
	By induction, for all $k$ we have
	$c_P(s_k)=c(s_k)2^{-k}$ and 
	$c_B(s_k)
	=\sum_{i\leq k} c_P(s_i)$.
	
	We have
	$c_B(s_k)=\sum_{i\leq k} c(s_i)2^{-i}
	\geq k 
	> \log\bar{c}(s_k)-1$.
	Since for all $s_k\prec s\prec s_{k+1}$,
	$\bar{c}(s)<2\bar{c}(s_k)$, we have that $\log\bar{c}(s)<\log\bar{c}(s_k)+1$, and therefore, for all $s$ we have $c'(s)\geq c_B(s)>\log\bar{c}(s)-2$.
\end{proof}

\begin{lemma}\label{lem_cws4conservative=>cws4Basic}
	If the Chooser has a computable winning strategy in the Betting game
	when playing against conservative Gamblers, 
	then the Chooser also has a computable winning strategy in the Betting game (that is, against any kind of Gambler).
\end{lemma}
\begin{proof}
	Let $G$ be any Gambler, and  
	let $G'$ be a Gambler that defines the same strategies as $G$ but with savings.
	Suppose that the Chooser has a computable winning strategy in the Betting game
	when playing against a conservative Gambler.
	Then there is some computably enumerable set of bounds $H'=\{h'_1,h'_2,\dots\}$ assigned to strategies defined by $G'$, $\bB'=\{B'_1,B'_2,\dots\}$,  such that the Chooser chooses a set 
	in the game against $G'$ with some size parameter $k$,
	that contains a sequence on which none of the strategies constructed by $G'$ achieve maximal capital larger than their bound.
	For every $n$ the $n$th strategy of $G$, $B_n$, 
	by Lemma \ref{slowly-but-surely}, achieves on that same sequence a maximal capital smaller than $h_n=2^{h'_n+2}$.
	Let $H=\{h_1,h_2,\dots\}$ and for all $n$, $h_n=2^{h'_n+2}$.
	The set of bounds $H$ is computably enumerable, and the Chooser wins in the game against $G$ with any size parameter $k$.
	That is, the Chooser has the same computable winning strategy in the Betting game
	when playing against any Gambler as the Chooser that plays against the conservative Gamblers, except that the capital bounds are exponentially larger.
\end{proof}

\begin{definition}\label{def_cwswr}
	The Chooser has a winning strategy \textit{with residue} for the Betting game on open sets, if for some set of capital bounds $H=\{h_1,h_2,\dots\}$ 
	and 
	every size parameter $k$
	there is some $n$ 
	such that 
	for every Gambler
	the Chooser wins the game when only the first $n$ of Gambler's betting strategies are considered. 
	That is,
	the chosen open set in the game
	has a subset, on which none of the first $n$ of Gamblers betting strategies
	achieve capital larger then their bounds when betting on any sequence in the subset.
	Additionally, this subset
	has measure that is larger than the sum $\sum_{i>n} 1/h_i$.
	
	Let $\bB_i=\{B^i_1,B^i_2,\dots \}$ denote the Gambler's betting strategies at the end of $i$th turn, and let $\bB^{1:n}_i=\{B^i_1,\dots,B^i_n \}$ the first $n$ of those strategies. 
	The Chooser's strategy is computable if $H$ is computable, $n$ is computable from $k$, and the chosen set in the $i$th turn is computable from Gambler's first $n$
	betting strategies at the end of the $(i-1)$th turn, $\bB^{1:n}_{i-1}$.
\end{definition}

\begin{lemma}\label{lem_cwswr=>cws}
	Suppose that for some set of capital bounds $H$ and the size parameter $k$,
	the Chooser chooses an open set $C$ in the Betting game 
	that has a subset $C'$ that contains sequences on which the first $n$ Gambler's betting strategies do not achieve capital larger than their capital bounds
	and the measure of $C'$ is larger than the sum $\sum_{i>n} 1/h_i$.
	Then $C$ contains a sequence on which none of the Gambler's betting strategies
	achieve capital larger then their capital bound.
\end{lemma}
\begin{proof}
	By assumption, the first $n$ betting strategies do not achieve capital larger than their bounds on any sequence in $C'$.
	The size of a set of sequences on which the $i$th betting strategy achieves capital larger than $h_i$ is at most $1/h_i$, and the size of the set of sequences on which at least one of the strategies after the $n$th
	achieves capital larger than its bound is then at most $\sum_{i>n} 1/h_i$.
	Since $\lambda(C')>\sum_{i>n} 1/h_i$, $C'$ must contain a sequence on which none of the strategies achieve capital larger than their bound.
\end{proof}

In the next chapter we will prove a key proposition
that will allow us to prove the main result.

\section{Proof of the main theorem}

\begin{proposition}[Key Proposition]\label{key_proposition}
	The Chooser has a computable winning strategy with residue in the Betting game against conservative Gamblers.
\end{proposition}

We will construct a strategy for the Chooser that chooses
clopen sets consisting of sequences that on a set of positions $I$ 
have the number of ones that has remainder $o$ when divided by natural number $m$ .
We will call such clopen sets \textit{modulo sets}.
The modulus $m$ remains the same throughout the Betting game, 
while set of positions $I$ and the remainder $o$ change.

%
\newcommand{\Mod}{\text{Mod}}
\begin{definition}[Modulo set]
	For any subset of positions $I$, modulus $m$ and remainder $o$,
	the set of strings whose number of ones on positions in $I$ modulo $m$
	is $o$ is denoted with $\Mod(I,m,o)$, that is,
	
	$\Mod(I,m,o)=\bigsqcup_{\{s:s\mod m=o\}}\{\sigma\in\{0,1\}^\infty:\sum_{p\in I}\sigma_p=s\}$.
\end{definition}

The modulo sets have the property that if the restriction has enough unrestricted positions in $I$, then the set of sequences consistent with the restriction is approximately independent of the modulo set (Corollary \ref{corr_Mr_indep}).

\begin{definition}\label{def_xi-approx}
	We will say that two reals $x,y$ are $\xi$-approximate, and write
	$x \approx_\xi y$ when
	$(1-\xi) y \leq x\leq \frac{1}{1-\xi} y$.
	
\end{definition}

\begin{definition}\label{def_xi-indep}
We say that a restriction $r$ is $\xi$-approximately independent of
a set of sequences $M$ when
$\lambda(M\cap\tilde r)\approx_\xi\lambda(M)\lambda(\tilde r)$
\end{definition}

\newcommand{\Ns}{\text{N*}}
\begin{definition}
	For any restriction $r$ and any set of positions $I$ we denote
	the number of positions in $I$ that are not restricted by $r$ with
	$\Ns(r,I)$, that is,
	
	$\Ns(r,I)=|\{p\in I : r_p=*\}|$.
\end{definition}

\begin{proposition}\label{MImo_approx_1/m}
	For any restriction $r$, modulus $m$ and $\xi<1$,
	if $\Ns(r,I)\geq (\frac{m}{\xi})^2$
	then for every remainder $o\in[0,m-1]$, $\lambda(\Mod(I,m,o)\;|\;\tilde r)\approx_\xi \frac{1}{m}$.
\end{proposition}
\begin{proof}
	Let $u$ be the number of positions in $I$ unrestricted by $r$ and let $j$ be the number of positions in $I$ that are restricted by $r$ to $1$.
	Let $S_i$ be the set of sequences that have value $1$ in $j+i$ positions in $I$.
	Clearly,
	\begin{equation}\label{interssum}
	  \sqs r \text{ intersects }S_i\text{ if and only if }i\in[0,u]. 	
	\end{equation}
	
	Furthermore, the measure of $S_i$, conditional on $\sqs r$, 
	is proportional to the number of ways we can restrict $u$ many positions
	so that $i$ many are restricted to value $1$ and the rest, $u-i$ many, to $0$.
	More precisely,
	\begin{equation}
			\lambda(S_i|\sqs r)={u \choose i}2^{-u}
	\end{equation}
 
    For the central binomial coefficient we can find an upper bound in \cite{PéterKomjáthMathOverflow}:
    \begin{equation}\label{cbc-upper-bound}
    	{2n\choose n}<\frac{4^n}{\sqrt{\pi n}}
    \end{equation}
 
 Denote the  measure of $S_i$, conditional on $\sqs r$, with $f(i)$.
 By \eqref{interssum}, for $i\not\in[0,u]$, $f(i)=0$. By properties of binomial coefficents, for $i\in[0,u]$:
 \begin{equation}\label{mid_sum_is_only_max}
 	f(i)\text{ is non-decreasing on }[0,\lfloor\frac{1}{2}u\rfloor]\text{ and
 	non-increasing on }[\lfloor\frac{1}{2}u\rfloor,u]
 \end{equation}
 .
 For the maximal value of $f$ we have:
    \begin{claim}\label{mid_sum_upper_bound}
    	$f(\lfloor\frac{1}{2}u\rfloor)<1/\sqrt{u}$
    \end{claim}
    \begin{proof}
    	If $u$ is even,
    	$f(\lfloor\frac{1}{2}u\rfloor)
    	=f(\frac{1}{2}u)
    	={u\choose u/2}2^{-u}$.
    	By \eqref{cbc-upper-bound} this is less than
    	$\frac{1}{\sqrt{\pi u}}
    	<1/\sqrt{u}$.
    	
    	On the other hand, if $u$ is odd,
    	$\lfloor\frac{1}{2}u\rfloor=(u-1)/2$.
        From definition of binomial coefficient, ${u\choose (u-1)/2}=\frac{1}{2}{u+1\choose (u+1)/2}$
        and we have
        $f((u-1)/2)
        ={u\choose (u-1)/2}2^{-u}
        =\frac{1}{2}{u+1\choose (u+1)/2}2^{-u}
        $.
        By \eqref{cbc-upper-bound} this is less than
        $\frac{1}{2}\frac{4^{(u+1)/2}}{\sqrt{\pi (u+1)}}2^{-u}
        =\frac{1}{\sqrt{\pi (u+1)}}
        <1/\sqrt{u}
        $
    \end{proof}
    
    \begin{claim}\label{single_max_f_even_odd_sum_diff}
    	Let $n_1,\dots,n_d,\dots,n_z$ be a sequence of $z$ non-negative numbers that is 
    	non-decreasing up to $d$th number, $n_d$, and non-increasing afterwards.
    	That is, $n_1\leq \dots \leq n_d\geq \dots \geq n_z$.
    	The difference between the sum of numbers with even indices
    	and the numbers with odd indices is at most $n_d$.
    \end{claim}
\begin{proof}
  We prove for the case when $d$ is odd and $z$ is even.
  The difference is then:
  $\sum_{i\in[1,z/2]}n_{2i-1}-n_{2i}\\
  =(\sum_{i\in[1,(d-1)/2]}n_{2i-1}-n_{2i})
  +n_d
  +(\sum_{i\in[(d+1)/2,z/2-1]}n_{2i+1}-n_{2i})-n_z$.
  Both sums are at most $0$, and so is $n_z$,
  therefore the difference is at most $n_d$.
  
  The other cases, when $d$ is even or $z$ is odd, can be reduced to the previous case by prepending $0$ at the beginning of the number sequence or appending $0$ at the end (or both). 
\end{proof}


	
	Denote with $M_o$ the modulo set defined on set of positions
	$I$ with modulus $m$ and remainder $o$, that is,
	$M_o=\Mod(I,m,o)$.
	Let
	$N_o$ denote the set of all $i\in[0,u]$ 
	such that $S_i$ intersects $M_o$, that is $(j+i)\mod m = o$.	
	We have that 
	$\lambda(M_o|\sqs r)=\sum_{i\in N_o} f(i)$.
	
	Let $o,o'$ be any two distinct remainders in $[0,m)$ and
	let $N$ be the sequence of numbers in $N_o\sqcup N_{o'}$, in increasing order. 
	For any two consecutive elements in $N$, one will be
	from $N_o$ and the other one from $N_{o'}$, and since $N$ is a sub-sequence
	of $[0,u]$, by \eqref{mid_sum_is_only_max} and claims \ref{mid_sum_upper_bound} and \ref{single_max_f_even_odd_sum_diff},
	the difference between the sum of $f$ over $N_{o}$ and the sum of $f$ over
	$N_{o'}$ is at most $1/\sqrt{u}$.
	This implies that there is some real $g$ such that for every
	$o$ in $[0,m)$ the sum of $f$ over $N_o$ is in $[g,g+1/\sqrt{u}]$.
	
	Since the modulo sets $M_0,\dots,M_{m-1}$ partition the set of sequences,
	we have $\sum_{o\in [0,m),i\in N_o}f(i)=1$.
	If $g<\frac{1}{m}-1/\sqrt{u}$ then the sum
	$\sum_{o\in [0,m),i\in N_o}f(i)$ would be strictly less than 1
	(a contradiction),
	and if $g>\frac{1}{m}$ this sum would be strictly more than 1 (a contradiction).
	Therefore $g\in[\frac{1}{m}-1/\sqrt{u},\frac{1}{m}]$, and we have that
	for every
	$o$ in $[0,m)$ the sum of $f$ over $N_o$ is between 
	$\frac{1}{m}-1/\sqrt{u}$ and $\frac{1}{m}+1/\sqrt{u}$.
	
	Since the sum of $f$ over $N_o$ is
	$\lambda(M_o|\sqs{r})$,
	and from the condition 
	$\Ns(r,I)\geq (\frac{m}{\xi})^2$,
	we have $1/\sqrt{u}\leq\frac{\xi}{m}$. 
	Therefore
	$\lambda(M_o|\sqs{r})$ is between
	$\frac{1}{m}(1-\xi)$ and $\frac{1}{m}(1+\xi)$.
	This implies the result as $1+\xi\leq 1/(1-\xi)$ for any $\xi<1$.
\end{proof}

\begin{corollary}\label{corr_M_size}
	Let $M$ be a modulo set defined on a set of positions $I$, with modulus $m$ and remainder $o$.
	That is, $M=\Mod(I,m,o)$.
	For any positive $\xi<1$,
	if $|I|\geq(\frac{m}{\xi})^2$ then
	$\lambda(M)\approx_\xi\frac{1}{m}$
\end{corollary}
\begin{proof}
	Let $r$ be the empty restriction, $r=\zeta$.
	Then $\tilde{r}$ is the set of all sequences and $\Ns(r,I)=|I|$.
	We have that $\Ns(r,I)\geq (\frac{m}{\xi})^2$.
	
	By Proposition \ref{MImo_approx_1/m},
	for any remainder $o$,
	we have
	$\lambda(\Mod(I,m,o))\approx_\xi \frac{1}{m}$.
\end{proof}

\begin{corollary}\label{corr_Mr_indep}
	Let $M$ be a modulo set defined on a set of positions $I$, with modulus $m$ and remainder $o$.
	That is, $M=\Mod(I,m,o)$.
	Let $r$ be a restriction that has more than $(\frac{m}{\xi})^2$  unrestricted positions in $I$, for some positive $\xi<1/2$.
	Then $\tilde r$ is $2\xi$-approximately independent of $M$.
\end{corollary}
\begin{proof}
	By Proposition \ref{MImo_approx_1/m} we have
	that $\lambda(M\cap\tilde{r})$ is between
	$\frac{\lambda(\tilde{r})}{m}(1-\xi)$ and
	$\frac{\lambda(\tilde{r})}{m}\frac{1}{(1-\xi)}$.
	From Corollary \ref{corr_M_size} we have that
	the measure of $M$ is between
	$\frac{1}{m}(1-\xi)$ and
	$\frac{1}{m}\frac{1}{(1-\xi)}$.
	This implies that $\lambda(M\cap\tilde{r})$ is
	at least $\lambda(\tilde{r})\lambda(M)(1-\xi)^2$
	and at most $\lambda(\tilde{r})\lambda(M)\frac{1}{(1-\xi)^2}$.
	Since $(1-2\xi)$ is less than $(1-\xi)^2$, the result follows.
\end{proof}

For a subset of positions $I$ we classify the restrictions according to the number of unrestricted positions in $I$.

\begin{definition}
	A restriction $r$ is \textit{$(I,\phi)$-chubby} if it has more than $\phi$ unrestricted positions in $I$, that is,
	$\Ns(r,I)\geq \phi$.
	
	A restriction is \textit{$(I,\phi)$-slim} if it is not $(I,\phi)$-chubby.
	
	We will say that an $(I,\phi)$-slim restriction \textit{restricts $I$ entirely} if it restricts all of the positions in $I$.
	
	A restriction is $(I,\phi)$-lean
	if it is \textit{$(I,\phi)$-slim} but does not restrict $I$ entirely.
	
	We will also say that the set of sequences $\tilde r$ is $(I,\phi)$-chubby (or slim, or lean) if
	the restriction $r$ is is $(I,\phi)$-chubby (or slim, or lean).
\end{definition}

Note that for a modulo set $M$ defined for positions $I$, modulus $m$ and a remainder $o$, if $\phi>(\frac{m}{\xi})^2$, then the $(I,\phi)$-chubby restrictions are $2\xi$-approximately independent of the modulo set $M$, and the restrictions that restrict $I$ entirely are either contained in $M$ or disjoint from it.

Let us look at the case when conservative Gambler defines bets of just one non-monotonic betting strategy.
For a set of sequences $X$ and a finite betting strategy $B$, we will consider the expectation of capital, with respect to $\lambda$, conditional on $X$, 
over the leaf-restrictions of a betting strategy,
or shortly, the expected capital of $B$ on $X$.
That is, denoting the leaf-strings of $B$ with $L$, 
the expected capital of $B$ on $X$ is
$\sum_{s \in L}\lambda(\sqs\rho(s)| X)c(s)$.
We can use the expected capital of $B$ on $X$ to upper bound 
the smallest capital over leaf-strings whose restriction intersects $X$.
 
If $B$ is conservative, we have that the
expected capital of $B$ on $X$ also upper bounds the
smallest maximum capital the finite betting strategy $B$ 
achieves when betting on sequences in $X$.
Namely, if the expected capital of $B$ for $X$ is $c$ then 
there is a leaf-string $s$ 
for which the maximum of capital is less than $c+2$ and
$\sqs\rho(s)$ intersects $X$.

Let $B,B'$ be finite conservative non-monotonic betting strategies and $B\to B'$. That is, $B'$ is obtained from $B$ by defining some additional bets that are also conservative.
Let $s$ be a leaf-string of $B$, and
$S$ the set of leaf-strings of $B'$ that have prefix $s$.

Suppose $\rho(s)$ is $(I,\phi)$-chubby.
If restrictions $\rho'(s')$ are also $(I,\phi)$-chubby for all $s'\in S$,
the modulo set $M$ is (approximately) independent of $\sqs\rho'(s')$ and
the expected capital on sequences in $M\cap\sqs\rho(s)$ remains (approximately)
$c(s)$.
In order to increase the expected capital on $M\cap\sqs\rho(s)$ to (approximately) 
$\frac{1}{1-d}c(s)$, the Gambler must define new bets so that
the measure of leaf-strings in $S$ that are assigned $(I,\phi)$-slim restrictions 
is at least $d\lambda(\sqs\rho(s))$.

On the other hand, suppose $\rho(s)$ restricts $I$ entirely, 
and $\sqs\rho(s)$ is a subset of $M$.
Regardless of how the additional bets are defined, 
the expected capital of $B'$ on $M\cap\rho(s)$ remains
$c(s)$.

We can see that if the betting strategy $B$ has no $(I,\phi)$-lean restrictions,
and $B'$ with new bets increases the expected capital on $M$ by a factor of
(approximately) $\frac{1}{1-d}$, then it must be that the measure of $(I,\phi)$-slim leaf-restrictions is increased by at least (approximately) $d$.

We can use essentially the same argument to show this is also true
for the expectation of expected capitals
when the Gambler has more than one betting strategy,
the $i$th one having probability of $2^{-i}$,
and the set of sequences contained in
$(I,\phi)$-lean restrictions can be non-empty, but small (Lemma \ref{KL_eta}).

\newcommand{\exc}[1]{\textit{earn}_{#1}} 
\begin{definition}
	Let $X$ be a set of sequences.
	For a betting strategy $B$, the expectation of capital, conditional on $X$, for the leaf-restrictions we will call in short the expected capital of $B$ \textit{on}
	$X$. That is, denoting the leaf-strings of $B$ with $S_B$, the expected capital of $B$ on $X$ is
	$\sum_{s\in S_B} c_B(s) \lambda(\tilde{\rho}(s)| X)$.
	
	Let  $\bB=\{B_1,B_2,\dots\}$ be
	a set of betting strategies that have finitely many bets defined
	and let the probability of the $i$th betting strategy in $\bB$ be $2^{-i}$.
	The expectation over the betting strategies in $\bB$ of 
	their expected capital on $X$,
	we call \textit{the expected earning for $\bB$ on $X$}
	and denote it with
	$\exc{\bB}(X)$.
	
	That is,
	
	$$\exc{\bB}(X)=
	\sum_{B_i\in \bB,s\in S_{B_i}}
	2^{-i}c_{B_i}(s)\lambda(\tilde{\rho_{B_i}}(s)| X)$$.
	
\end{definition}
In the next four lemmas (\ref{lem_exc_of_Omega}, \ref{lem_exc_of_subset},
\ref{lem_exc_part_of_X}, \ref{lo_cap_sub}) we prove some properties of the expected earning that we will use later.
\begin{lemma}\label{lem_exc_of_Omega} 
	For any set of finite betting strategies $\bB$ the expected earning on the entire set of sequences is less than $1$.
\end{lemma}
\begin{proof}
	For strategy $B$, the expected capital on the set of all sequences is 
	\\
	$\sum_{s\in S_{B}}c_{B}(s)\lambda(\tilde{\rho}(s))
	=\sum_{s\in S_{B}}\mu_B(s)=1$.
	\\
	The expected earning is the expectation of expected capitals,
	\\
	$\exc{\bB}(\{0,1\}^\infty)=
	\sum_{B_i\in \bB}
	\sum_{s\in S_{B_i}}2^{-i}$.
\end{proof}
\begin{lemma}\label{lem_exc_of_subset} 
	For any  set of finite betting strategies $\bB$, and any set of sequences $X$ and its subset $X'$,
	$\exc{\bB}(X')\leq \frac{\lambda(X)}{\lambda(X')}\exc{\bB}(X)$.
\end{lemma}
\begin{proof}
	We have
	$\exc{\bB}(X')=\frac{1}{\lambda(X')}
	\sum_{B_i\in \bB}
	\sum_{s\in S_{B_i}}2^{-i}c_{B_i}(s)\lambda(\tilde{\rho}(s)\cap X')\\
	\leq
	\frac{1}{\lambda(X')}
	\sum_{B_i\in \bB}
	\sum_{s\in S_{B_i}}2^{-i}c_{B_i}(s)\lambda(\tilde{\rho}(s)\cap X)
	=
	\frac{\lambda(X)}{\lambda(X')}\exc{\bB}(X)
	$
\end{proof}
\begin{lemma}\label{lem_exc_part_of_X}
	For every set of sequences $X$, any finite partition of $X$ and any set of finite betting strategies $\bB$ there is a part in the partition such that the expected earning of $\bB$ on the part is less than the expected earning of $\bB$ on the set $X$.
\end{lemma}
\begin{proof}
	The result immediately follows from the law of total expectation.
\end{proof}

\begin{lemma}\label{lo_cap_sub}
	Let $\bB=\{B_1,B_2,\dots\}$ be a countable set of finite betting strategies. 
	Every set of sequences $X$ for any $d$ smaller than the measure of $X$ contains a subset $Y$ of measure more than $d$ such that
	for every strategy $B_i\in\bB$ and every sequence in $Y$ the leaf-restriction that contains the sequence has capital less than 
	$2^i\frac{\lambda(X)}{\lambda(X)-d}\exc{\bB}(X)$.
\end{lemma}
\begin{proof}
	Let $q=\frac{\lambda(X)}{\lambda(X)-d}$ and let $Z$ be the set of sequences in $X$ 
	for which
	there is some
	betting strategy $B_i\in\bB$ whose leaf-restriction that contains 
	the sequence has capital strictly larger than $q2^i\exc{\bB}(X)$.
	
	We have that $\exc{\bB}(Z)\geq q\exc{\bB}(X)$. 
	By Lemma \ref{lem_exc_of_subset},
	$\exc{\bB}(Z)\leq \frac{\lambda(X)}{\lambda(Z)}\exc{\bB}(X)$,
	and we have
	$\lambda(Z)\leq\frac{\lambda(X)}{q}=\lambda(X)-d$.
	
	Let $Y=X\setminus Z$.
	The measure of $Y$ is larger than $d$,
	and for every sequence in $Y$
	and every strategy $B_i\in\bB$ the leaf-restriction that contains the sequence has capital less than $q2^i\exc{\bB}(X)$.
		
\end{proof}

\begin{definition}
	A set where we allow multiple instances of elements is called a
	\textit{multi-set}. 
	The number of instances for an element in a multi-set is called 
	the multiplicity of the element in the multi-set.
	A set is then a multi-set where each element has multiplicity 1.
	We will be considering only multi-sets of finite restrictions.
	
	The sum of measures of sets of sequences consistent with restrictions
	in the multi-set $R$ is denoted with $\lplus(R)$ 
	and called the \textit{sum-size} of $R$, that is,
	
	$\lplus(R)=\sum_{r\in R}\lambda(\sqs{r})$.
	The set of sequences that are consistent with any restriction in $R$ is denoted with $\sqs R$.
	
	A union of two multi-sets is called a join. 
	The multiplicity of an element in the join of two multi-sets is the
	sum of multiplicities of the element in the two multi-sets.
\end{definition}
For example, let $R$ be the join of leaf-restrictions of $n$ finite non-monotonic betting strategies. Then  $\lplus(R)=n$ and $\lambda(\sqs R)=1$.
\begin{definition}
	Let $B=(\rho,\mu)$ be a finite Kolmogorov-Loveland betting strategy,
	and $B'=(\rho',\mu')$ a Kolmogorov-Loveland betting strategy obtained from $B$
	by defining finitely many new bets, that is,
	$B\to B'$.	
	Let $s,s'$ be leaf-strings of $B,B'$ such that $s$ is a prefix of $s'$.
	Denote with $r,r'$ the restrictions $\rho(s),\rho'(s')$.
	
	If for some set of positions $I$ and number $\phi$ the restriction $r$ is $(I,\phi)$-chubby and the restriction $r'$ is $(I,\phi)$-slim
	we will say that $r'$ \textit{was $(I,\phi)$-slimmed down by new bets of $B'$}.
\end{definition}

\begin{lemma}\label{KL_eta}
	Let $M=\Mod(I,m,o)$ be some modulo set and 
	let $\phi>m^2$, 
	$\xi=m/\sqrt{\phi}$.
	
	Let $\bB$ be a finite set of finite Kolmogorov-Loveland betting strategies,
	and let $\bB'$ be a set of Kolmogorov-Loveland betting strategies obtained from $\bB$
	by defining finitely many new bets , that is,
	$\bB\to \bB'$.

	Denote with $\Theta$
	the multi-set of $(I,\phi)$-lean leaf-restrictions of strategies in $\bB$.

	Denote with $\Delta$ the multi-set of
	restrictions that were slimmed down by newly defined bets of strategies in $\bB'$.
	
	Denote with $M'$ the sequences in $M$ that are not consistent with any restriction in $\Delta$ or $\Theta$. 
	
	The expected earning for $\bB'$ on $M'$, $\exc{\bB'}(M')$
	is at most
	
	$\frac{1}{(1-2\xi)(1-\lambda(\tilde\Theta|M))-\lplus(\Delta)}\exc{\bB}(M)$.
\end{lemma}
\begin{proof} Let 
	$\bB=\{B_1,\dots,B_n\}$,
	$\bB'=\{B'_1,\dots,B'_n\}$.
	Since $\bB\to\bB'$, we also have $B_i\to B'_i$.
	For $i\leq|\bB|$, denote the
	restriction functions of $B_i,B'_i$ with $\rho_i,\rho'_i$, 
	their mass functions with $\mu_i,\mu'_i$ and capitals with $c_i,c'_i$.
	
	Let 
	$S_i$ denote the leaf-strings of the betting strategy $B_i$ and let
	$S'_i$ denote the leaf-strings of the betting strategy $B'_i$.

Let 
$\Psi_i$ be the strings in $S_i$ that are assigned by $\rho_i$ restrictions
that restrict $I$ entirely.
Let 
$\Psi'_i$ be the strings in $S'_i$ that have a prefix in $\Psi_i$.
\begin{claim}\label{PsiPsisum}
	$
	\sum_{s' \in \Psi'_i}c'_i(s')\lambda(\tilde\rho'(s')\cap M)
	=\sum_{s \in \Psi_i}c_i(s)\lambda(\tilde\rho(s)\cap M)$.
\end{claim}
\begin{proof}
	A restriction $\rho(s), s\in\Psi_{i}$ restricts all positions in $I$ and $\tilde\rho(s)$
	is either contained in $M$ or disjoint from it.
	This is also true for any restriction $\rho'(s'), s'\in\Psi'_{i}$ 
	and both values 
	$\lambda(M|\tilde\rho'(s'))$ and $\lambda(M|\tilde\rho(s))$ are the same 
	(they are either 0 or 1)
	when
	$s\preceq s'$. 	
	Therefore,
	
		$
		\sum_{s' \in \Psi'_i}c'_i(s')\lambda(\tilde\rho'(s')\cap M)\\
		=\sum_{s'\in \Psi'_i}\mu'_i(s')\lambda(M|\tilde\rho'(s'))\\
		=\sum_{s\in\Psi_i}\lambda(M|\tilde\rho(s))\sum_{s'\in \Psi'_i,s\preceq s'}\mu'_i(s')\\
		=\sum_{s\in \Psi_i}\mu_i(s)\lambda(M|\tilde\rho(s))
		=\sum_{s \in \Psi_i}c_i(s)\lambda(\tilde\rho(s)\cap M)$.
\end{proof}
	
	Let
	$\Phi_i, \Phi'_i$ be the strings in $S_i,S'_i$ that are assigned $(I,\phi)$-chubby restrictions by $\rho_i,\rho'_i$, respectfully.
	\begin{claim}\label{PhiPhisum}
		$\sum_{s'\in \Phi'_i}c'_i(s')\lambda(\tilde\rho'(s')\cap M)
		\leq
		\frac{1}{(1-2\xi)}\sum_{s\in\Phi_i}c_i(s)\lambda(\tilde\rho(s)\cap M) 
		$.
	\end{claim}
\begin{proof}
		A string $s'\in\Phi'_i$ has a prefix $s\in\Phi_i$,
		and
		by Proposition \ref{MImo_approx_1/m}, both
		values $\lambda(M|\tilde\rho'(s'))$ and $\lambda(M|\tilde\rho(s))$ are $\xi$-approximately $\frac{1}{m}$.
		At the most extreme, one value is $1/(1-\xi)$ times larger than $\frac{1}{m}$,
		and the other one is $1-\xi$ times smaller, implying the two values are
		$2\xi-\xi^2$ (and therefore also $2\xi$) approximate. We have:
		
		$\sum_{s'\in \Phi'_i}c'_i(s')\lambda(\tilde\rho'(s')\cap M)\\
		=
		\sum_{s'\in \Phi'_i}\mu'_i(s')\lambda(M|\tilde\rho'(s'))\\
		\approx_{2\xi}
		\sum_{s\in\Phi_i}\sum_{s'\in \Phi'_i,s\preceq s'}\mu'_i(s')\lambda(M|\tilde\rho(s))\\
		=
		\sum_{s\in\Phi_i}\lambda(M|\tilde\rho(s))\sum_{s'\in \Phi'_i,s\preceq s'}\mu'_i(s')\\
		\leq
		\sum_{s\in\Phi_i}\mu_i(s)\lambda(M|\tilde\rho(s))\\
		=
		\sum_{s\in\Phi_i}c_i(s)\lambda(\tilde\rho(s)\cap M) 
		$.
\end{proof}

By claims \ref{PsiPsisum},\ref{PhiPhisum}, for all $i\leq|\bB|$, 
\begin{equation}\label{sumPhiPsi}		
	\sum_{s'\in \Phi'_i\sqcup\Psi'_i}c'_i(s')\lambda(\tilde\rho'(s')\cap M)
	\leq
	\frac{1}{(1-2\xi)}\sum_{s\in \Phi_i\sqcup\Psi_i}
	c_i(s)\lambda(\tilde\rho(s)\cap M)
\end{equation}.

Let
$\Delta_{i}$ denote the strings in $S'_i$ that are assigned by $\rho'_i$ a restriction that was slimmed down by new bets.

Let 
$\Theta_i$ be the leaf-strings that are assigned $(I,\phi)$-lean restrictions
by $\rho_i$.
Let
$\Theta'_i$ be the leaf-strings of $B'_i$ that have a prefix in $\Theta_i$.

Since the restrictions are either $(I,\phi)$-chubby, $(I,\phi)$-lean or restrict $I$ entirely, we have
$S_i=\Phi_{i}\sqcup\Theta_i\sqcup\Psi_{i}$.

A leaf-string in $\Phi_i$ has successors in $S'_i$ that are assigned by $\rho'_i$
restrictions that are either $(I,\phi)$-chubby or were slimmed down by new bets,
therefore
$S'_i=\Phi'_{i}\sqcup\Delta_i\sqcup\Theta'_i\sqcup\Psi'_{i}$
.

The multi-set $\Delta$ is the join of $\bigcup_{s'\in\Delta_{i}}\rho'(s')$ over all $i\leq|\bB|$,
the multi-set $\Theta$ is the join of $\bigcup_{s\in\Theta_i}\rho_i(s)$ over all $i\leq|\bB|$
and
$M'=M\setminus(\tilde{\Delta}\cup\tilde\Theta)$.

The sequences in $M'$ are consistent only with chubby leaf-restrictions of $B'_i$ and the leaf-restrictions of $B_i$ that restrict $I$ entirely,
that is, for any leaf-string $s'$ in $\Delta_i\cup\Theta'_i$, 
$\lambda(\tilde{\rho}'(s')\cap M')=0$, and we have

$\lambda(M')\exc{\bB'}(M')
=
\sum_{i\leq |\bB|}
\sum_{s'\in \Phi'_i\sqcup\Psi'_i}2^{-i}c'_i(s')\lambda(\tilde\rho'(s')\cap M')
$

Since $M'\subseteq M$,

$\lambda(M')\exc{\bB'}(M')
\leq
\sum_{i\leq |\bB|}
\sum_{s'\in \Phi'_i\sqcup\Psi'_i}2^{-i}c'_i(s')\lambda(\tilde\rho'(s')\cap M)
$
\\ 
By \eqref{sumPhiPsi}:

$
\sum_{s'\in \Phi'_i\sqcup\Psi'_i}c'_i(s')\lambda(\tilde\rho'(s')\cap M)
\leq
\frac{1}{1-2\xi}
\sum_{s\in \Phi_i\sqcup\Psi_i}c_i(s)\lambda(\tilde\rho(s)\cap M)$
\\	Since $\Phi_i\sqcup\Psi_i$ is a subset of all leaf-strings of strategy $B_i$ we have

$\sum_{s\in \Phi_i\sqcup\Psi_i}c_i(s)\lambda(\tilde\rho(s)\cap M)
\leq
\sum_{s\in S_i}c_i(s)\lambda(\tilde\rho(s)\cap M)
$
\\
Summing over $i$ we get:

$\sum_{i\leq |\bB|}\sum_{s'\in \Phi'_i\sqcup\Psi'_i}2^{-i}c'_i(s')\lambda(\tilde\rho'(s')\cap M)\\
\leq
\frac{1}{1-2\xi}
\sum_{i\leq |\bB|}\sum_{s\in S_i}2^{-i}c_i(s)\lambda(\tilde\rho(s)\cap M)
\\
=
\frac{1}{1-2\xi}\lambda(M)\exc{\bB}(M)$.	
Therefore

$$\lambda(M')\exc{\bB'}(M')
\leq
\frac{1}{1-2\xi}\lambda(M)
\exc{\bB}(M)
$$.

Since $\lambda(M')\geq\lambda(M)-\lambda(M\cap\tilde\Theta)-\lambda(M\cap\tilde{\Delta})$,
we have

\begin{equation}\label{eq_M_M'_measure}
\frac{\lambda(M)}{\lambda(M')}
\leq
\frac{1}{1-\lambda(\tilde\Theta|M)-\lambda(\tilde{\Delta}|M)}
\end{equation}

Finally,
$\exc{\bB'}(M')
\leq
\frac{1}{(1-2\xi)}\frac{1}{1-\lambda(\tilde\Theta|M)-\lambda(\tilde{\Delta}|M)}
\exc{\bB}(M)$.
The only thing left to prove is that
\begin{equation}\label{eq_D|M_D}
\lambda(\tilde{\Delta}|M)\leq\lplus(\Delta)/(1-2\xi).
\end{equation}

Suppose that for some string $v$ that is a prefix of a slimmed-down leaf-string, 
the restriction $\rho'_i(v)$
has exactly $\phi$ unrestricted positions.
This restriction is still chubby, and if the strategy then bets on a position
in $I$, both of its immediate successors $\rho'(v0),\rho'(v1)$ are slim.
The sum over leaf-strings $s'$ of $B'_i$ that extend $v$ 
of $\lambda(\rho'_i(s')\cap M)$
is equal to $\lambda(\rho'_i(v)\cap M)$.

By Corollary \ref{corr_Mr_indep},
the restriction $\rho'_i(v)$ is 
$2\xi$-independent of $M$.
Let $D_i$ denote the set of sequences consistent with
the leaf-restrictions that were slimmed-down by new bets of strategy $B'_i$.
That is, $D_i=\bigsqcup_{s'\in\Delta_i}\tilde\rho'(s')$.
The measure of $D_i$ conditional on $M$
is at most $\lambda(D_i)/(1-2\xi)$.
Since $\tilde{\Delta}=\bigcup_{i\leq|\bB|} D_i$,
we have that 
$\lambda(\tilde{\Delta}|M)\leq\sum_{i\leq|\bB|}\lambda(D_i|M)$
which is less than $\lplus(\Delta)/(1-2\xi)$.

\end{proof}

In Lemma \ref{KL_eta}, we have shown that the Gambler,
starting a turn of the betting game with a finite set of Komlogorov-Loveland betting strategies $\bB$,
in order to increase the expected earning on sequences 
in a modulo set defined on the set of positions $I$ with modulus $m<\sqrt{\phi}$,
has to define new bets for the strategies that slim down a large multi-set of $(I,\phi)$-chubby leaf-restrictions of strategies in $\bB$,
under the condition that the multi-set of $(I,\phi)$-lean 
leaf-restrictions of strategies in $\bB$ is small.

Suppose that, to the contrary, the multi-set of $(I,\phi)$-lean 
leaf-restrictions of strategies in $\bB$ is not small.
We will show that, if $\phi$ is small compared to $I$,
we can find a large subset of $I$, $I'$, such that
most of those $(I,\phi)$-lean leaf-restrictions 
restrict $I'$ entirely.

\newcommand{\lf}{\lfloor}
\newcommand{\rf}{\rfloor}
\newcommand{\lc}{\lceil}
\newcommand{\rc}{\rceil}
\begin{lemma}\label{lem_slim_to_restricted}
	For any multi-set $R$ of $(I,\phi)$-slim restrictions
	and any positive $q<1$, 
	there is some
	$I'\subseteq I$ such that 
	$|I'|\geq q|I|$ and
	the sum-size of restrictions in $R$
	that restrict $I'$ is at least 
	$(1-\phi q')\lplus(R)$, where
	$q'=1/(\lf \frac{|I|}{\lc q|I| \rc} \rf)$ .
\end{lemma}
\begin{proof}
	Let $\ell=\lc q|I| \rc$, 
	$n=\lfloor\frac{|I|}{\ell}\rfloor$,
	and let $I_1,\dots, I_n$ be disjoint consecutive subsets of $I$ with $|I_i|=\ell$.

	For a set of positions $J$, let $L(J)$ denote
	the sum over all restrictions in the multiset $R$ of
	the product of 
	the number of unrestricted positions in $J$
	and the measure of the restriction.
	That is,

	$L(J)=\sum_{r\in R}\Ns(r,J)\lambda(\sqs{r})$.
	\\
	Suppose that the proposition is not true and
	for every subset of $I$ with $\ell$ elements,
	the sum-size of restrictions that have one (or more)
	unrestricted positions in the subset is more than 
	$\phi q'\lplus(R)$.
	This is also true for every  $I'\in\{I_1\dots I_n\}$.

	The sum-size of restrictions that have one (or more)
	positions in $I'$ unrestricted
	is a lower bound for $L(I')$.
	Since $\{I_1\dots I_n\}$ are disjoint subsets of $I$,
	$L(I)\geq\sum_{i\in[1,n]}L(I_i)$, and we have that
	$L(I)\geq n\phi q'\lplus(R)$.
	
	Since $q'=\frac{1}{n}$, 
	$L(I)\geq \phi\lplus(R)$.
	This implies that there is some restriction in $R$
	with at least $\phi$ unrestricted positions, and
	$R$ therefore contains an $(I,\phi)$-chubby restriction,
	contrary to the assumption.
	

\end{proof}

It will be easier to use the following corollary of the previous lemma as it does not have rounding.
\begin{corollary}\label{lem_slim_to_restricted_corr}
	For any multi-set $R$ of $(I,\phi)$-slim restrictions
	and any $q$ between $\frac{3}{|I|}$ and $\frac{1}{4}$
	there is
	some $I'\subseteq I$ such that 
	$|I'|\geq q{|I|}$ and
	the sum-size of restrictions in $R$
	that restrict $I'$ is at least 
	$(1-2q\phi)\lplus(R)$.
\end{corollary}
\begin{proof}
	It is enough to show that 
	if 
	$q\leq 1/4$ and $|I|\geq 3/q$
	then 
	$1/\lf \frac{|I|}{\lc q|I| \rc} \rf=q'\leq 2q$.
	The result then follows from Lemma \ref{lem_slim_to_restricted}.
	
	We have:
	\\
	$1/\lf \frac{|I|}{\lc q|I| \rc} \rf
	<
	1/(\frac{|I|}{q|I|+1}-1)
	=\frac{q|I|+1}{|I|-(q|I|+1)}
	$. 
	We will find a bound on $|I|$ and $q$ so that this last term, 
	$\frac{q|I|+1}{|I|-(q|I|+1)}$, is less than $2q$.
	
	We have:
	\\
	$\frac{q|I|+1}{|I|-(q|I|+1)}\leq 2q
	\iff
	q|I|+1\leq 2q|I|-2q^2|I|-2q
	\iff
	1+2q\leq q|I|-2q^2|I|\\
	\iff
	\frac{1}{q}\frac{1+2q}{1-2q}\leq |I|
	$.
	
	It is easy to see that this last inequality is true for any $|I|\geq 3/q$
	and $q\leq 1/4$.
\end{proof}

Suppose that the sum-size of $(I,\phi)$-lean leaf-restrictions of a finite set of finite Kolmogorov-Loveland betting strategies is above some desired bound $\delta$.
We can use Corollary \ref{lem_slim_to_restricted_corr}, setting $q=\delta/\phi$,
to find a set of positions $I'\subseteq I$ where
most of the $(I,\phi)$-lean leaf-restrictions restrict $I'$ entirely.
The sum-size of the leaf-restrictions that restrict $I'$ entirely
is larger than the sum-size of the leaf-restrictions that restrict $I$ entirely
by at least $\delta'=\delta(1-2\delta)$.

It could be that a lot of the $(I,\phi)$-chubby leaf-restrictions are
$(I',\phi)$-lean.
In this case we can again use Corollary \ref{lem_slim_to_restricted_corr}
to find $I''\subset I'$ where most of the $(I',\phi)$-lean leaf-restrictions restrict $I''$ entirely,
and so on. However, this cannot go on forever as
each time we find a new subset, the sum-size of the leaf-restrictions that restrict
the entire subset is incremented by $\delta'$ and
 at some point we would find a set of positions that is entirely restricted by all of the leaf-restrictions.

\begin{lemma}\label{smalling_slim_larging_restricted}
	Let $R$ be a
	multi-set of finite restrictions,
	$I$ a set of positions,
	and $\phi$ a bound on the number of unrestricted positions with $\phi\geq 2$.
	Let $x$ be the sum-size of restrictions in $R$ that restrict $I$ entirely.
	Let $\delta<1/2$ and $\delta'=\delta(1-2\delta)$.
	Let $g$ be the smallest integer such that
	$g\delta'+x\geq\lplus(R)$.
	Let $q=\delta/\phi$.
	
	If
	$|I|$ is larger than $(1/q)^{g+2}$
	then there is some positive integer $k\leq g$ and $I'\subseteq I$
	such that
	$|I'|\geq q^k|I|$
	and the sum-size of restrictions in $R$ that are $(I',\phi)$-lean is at most $\delta$,
	while the sum-size of restrictions in $R$ that restrict $I'$ entirely
	is at least $x+k\delta'$. 
\end{lemma}
\begin{proof}

	Let 
	$\Theta$ be the multi-set of $(I,\phi)$-lean restrictions from $R$.
	If  $\Theta$ already
	has sum-size smaller than $\delta$,
	then $I'=I$.
	On the other hand,
	if $\Theta$ has sum-size larger than $\delta$,
    since $q=\delta/\phi\leq 1/4$,
    $|I|\geq(1/q)^2$,
    and
    $\frac{3}{|I|}\leq 3q^2\leq q$,
    we can use Corollary \ref{lem_slim_to_restricted_corr}
    to find some $I'$ with $|I'|\geq q|I|$,
	such that the sum-size of restrictions in $\Theta$
	that restrict $I'$ is at least
	$(1-2\delta)\lplus(\Theta)$  which is larger than $\delta'$.
	Let $x'$ be the sum-size of restrictions that 
	restrict $I'$ entirely.
	Since the restrictions that restrict $I$ entirely, also restrict $I'$ entirely,
	$x'\geq x+\delta'$.
	
	Some of the $(I,\phi)$-chubby restrictions might be
	$(I',\phi)$-slim. If
	the multi-set of $(I',\phi)$-lean restrictions from $R$ 
	has sum-size larger than $\delta$
	we can repeat the same argument on $I'$. 
	This adds another $\delta'$
	to the sum-size of restrictions that restrict the set $I''$
	with size larger than $q^2|I|$.
	We can keep repeating this until we find a subset with
	less than $\delta$ slim restrictions that do not restrict it. 
	We are guaranteed that we will eventually find such subset,
	as after applying the argument $g$ many times, we find
	a set of positions such that 
	the sum-size of restrictions that restrict it is
	$\lplus(R)$ and therefore the sum-size of the rest of the slim restrictions
	is zero.
\end{proof}

%
%

Suppose that for some finite set of finite Kolmogorov-Loveland betting strategies $\bB$, set of positions $I$ and the bound on the number of unrestricted positions $\phi$, the set of sequences consistent with $(I,\phi)$-lean leaf-restrictions of strategies in $\bB$
is small.
Then for any modulus $m$, the modulo sets defined on $I$ with remainders
$0,\dots,m-1$ partition the set of infinite binary sequences, and
there must be some remainder $o$, such that the modulo set $\Mod(I,m,o)$ has both low earning
and small intersection with the set of sequences consistent with $(I,\phi)$-lean leaf-restrictions.

\begin{lemma}\label{good_mod}
	For any finite set of finite Kolmogorov-Loveland betting strategies $\bB$,
	any set of sequences $\Theta$ with measure less than $\delta$, 
	every modulus $m$ and bound $c>1$, 
	there is a remainder $o$
	such that 
	the expected earning for $\bB$ on 
	the modulo set $M=\Mod(I,m,o)$ 
	is less than $1/(1-\frac{1}{c})$
	and the measure of the intersection of $M$ with $\Theta$ is less
	than $c\delta\lambda(M)$. That is,
	
	$\exc{\bB}(M)\leq 1/(1-\frac{1}{c})$,
	and
	
	$\lambda(\Theta|M)\leq c\delta$
\end{lemma}
\begin{proof}
	The family of modulo sets with different remainders partitions the set of sequences.
	The sum of measures of modulo sets for which the measure of $\Theta$,
	conditional on the modulo set, is larger than $c\lambda(\Theta)$ is 
	at most $\frac{1}{c}$.
	Let $X$ denote the union of modulo sets for which the measure of $\Theta$,
	conditional on the modulo set is less than
	$c\delta$. 
	We have that the measure of $X$ is at least $1-\frac{1}{c}$.
	By lemmas \ref{lem_exc_of_Omega}, \ref{lem_exc_of_subset} we have that
	$\exc{\bB}(X)\leq 1/(1-\frac{1}{c})$, and by Lemma \ref{lem_exc_part_of_X},
	for at least one of these modulo sets the expected earning is
	smaller than  $1/(1-\frac{1}{c})$.
\end{proof}

We will construct a computable winning strategy with residue for the Chooser in the Betting game against conservative Gamblers called the Modulo Chooser.
We fix the set of capital bounds $H=\{h_1,h_2,\dots\}$ so that
$h_i=2^{i+5}$.
Given the size parameter $k$, we will pick a large enough modulus $m$ and large enough number of strategies $n$.
We will show that the Modulo Chooser, throughout the entire game chooses an open set
of measure less than $2^{-k}$, and the measure of chosen sequences on which none of the first $n$ of Gambler's betting strategies achieve capital larger their respective capital bounds is more than $2^{-n}$, satisfying Definition \ref{def_cwswr}.

At the beginning of the first turn of the Betting game,
the Chooser picks large enough $\phi,I_1$, 
and chooses the modulo set $M_1=\Mod(I_1,m,0)$.
Note that the leaf-restriction of the initial betting strategy
is $(I_1,\phi)$-chubby since it is the empty restriction and does not restrict any positions, also the set of $(I_1,\phi)$-lean leaf-restrictions is empty.
Therefore, before the Gambler has defined any bets, the expected earning on $M_1$ is less than $1$.

We can use Lemma \ref{KL_eta}, 
to show that the Gambler 
must define some additional bets for the betting strategies 
that bet on at least $|I_1|-\phi+1$ positions in $I_1$ 
to slim down the empty restriction,
and furthermore,
the sum-size of the multi-set of $(I_1,\phi)$-slim leaf-restrictions  
must be large.

If this does not happen, then by Lemma \ref{lo_cap_sub} there is a subset of $M_1$
of measure larger than $2^{-n}$
on which none of the Gambler's betting strategies achieve capital larger than their bound,
and already with the first chosen set the Chooser wins the game.

On the other hand, if the Gambler does define new bets so that the 
the multi-set of $(I_1,\phi)$-slim leaf-restrictions is large,
then
by Lemma \ref{smalling_slim_larging_restricted} the Chooser can find
a large $I_2\subseteq I_1$ such that
the sum-size of $(I_2,\phi)$-lean leaf-restrictions is small enough,
and the sum-size of leaf-restrictions 
that restrict $I_2$ entirely is larger than some amount $d$.

Next, by Lemma \ref{good_mod} the Chooser can find some remainder $o_2$,
so that the modulo set $M_2=\Mod(I_2,m,o_2)$ has small intersection with sequences consistent with $(I_2,\phi)$-lean leaf-restrictions and has low expected earning.
Again, low earning implies that there are sequences in $M_2$ on which the Gambler's betting strategies at the beginning of the second turn do not achieve capital higher than their bounds.
Then, again by Lemma \ref{KL_eta}, the Gambler must define some additional bets
so that the sum-size of the multi-set of $(I_1,\phi)$-slim leaf-restrictions  
is large, and so on.

After the Chooser chooses the $i$th (nonempty) set we are guaranteed that 
either the Chooser wins the game or, 
when the Gambler has defined additional bets,
we find some large set of positions on which 
the sum-size of the leaf-restrictions that restrict this entire set is at
least $i\cdot d$.

The game cannot go on forever as after the $(n/d)$th set was chosen there would be
some set of positions that is entirely restricted by all of the leaf-restrictions,
implying that the Gambler cannot increase earning on the 
modulo set chosen in the $(n/d+1)$st turn 
and the Chooser certainly wins the game.
Since $I_1$ was chosen large enough, and all of the 
$I_1\supseteq I_2\supseteq\dots\supseteq I_{n/d+1}$ are large compared to the modulus $m$, all of the chosen modulo sets have measure approximately $1/m$.
Since $m$ was chosen large enough the total size of the chosen sets is 
smaller than $2^{-k}$ (the bound given by the size parameter $k$).

We now give a formal definition of the Modulo Chooser.
\begin{definition}
	We define a strategy for the Chooser in the Betting Game on open sets against conservative Gamblers called \textit{the Modulo Chooser}.

    Let the capital parameter be $H=\{h_1,h_2,\dots\}$, with $h_i=2^{i+5}$, and let $k$ be the size parameter.

	Let modulus $m$ be $2^{2(k+4)}$.
	Let the number of Gambler's strategies that are considered be $n=4+\log m$.
	Let the bound on the number of unrestricted positions be
	$\phi=64m^2$.
	Let $\ell=(4\phi)^{8n+3}$.
	
	The Modulo Chooser will be choosing modulo sets defined on some subset of
	the first $\ell$ positions with modulus $m$ and some remainder.
	In the first turn, the chosen modulo set is defined on entire interval 
	of positions $I_1=[1,\ell]$ with modulus $m$ and remainder $o_1=0$, we denote this modulo set $M_1$.
	
	Let $j_i$ denote the turn when the $i$th modulo set, $M_i=\Mod(I_i,m,o_i)$, is chosen. 
	Suppose that in turn $j_i$ and subsequent turns the Gambler
	defines new bets, and by the end of turn $t$   
	the sum-size of the restrictions that were $(I_i,\phi)$-slimmed down 
	by the new bets for the first $n$ strategies becomes more than $3/8$.
	If, and only if, this happens the Modulo Chooser will choose the next modulo set, $M_{i+1}$, at the beginning of turn $j_{i+1}=t+1$, otherwise $j_{i+1}$ is undefined, and the Modulo Chooser does not choose any more (nonempty) sets.
	When choosing the $(i+1)$st modulo set, the Modulo Chooser finds a subset of the first $\ell$ positions, $I_{i+1}$,
	with properties:
	\begin{enumerate}[label=\textbf{I.\arabic*}]
		\item\label{atomic_large}
			Let $\Psi_i$ denote the multiset of leaf restrictions of strategies in $\bB^{1:n}_{t}$
			that restrict $I_{i+1}$ entirely.
			Let $z_i=8\lplus(\Psi_i)$.
			The sum-size of $\Psi_i$ is more than $i/8$ and the number of positions in $I_{i+1}$ is more than $(\frac{1}{4\phi})^{z_i}\ell$.

		\item\label{lean_small}
			Let $\Theta_i$ denote 
			the multiset of leaf restrictions of strategies in $\bB^{1:n}_{t}$
			that are $(I_{i+1},\phi)$-lean.
			The sum-size of $\Theta_i$ is less than $1/4$.
	\end{enumerate}
	The Modulo Chooser then chooses a modulo set $M_{i+1}$ defined on the interval $I_{i+1}$,
	with modulo $m$ and a remainder $o_{i+1}$ with properties:
	\begin{enumerate}[label=\textbf{M.\arabic*}]
		\item\label{M_lean_small}
		The size, conditional on $M_{i+1}$,
		of the set of sequences contained in an $(I_{i+1},\phi)$-lean leaf-restriction of any of the Gambler's betting strategies, after the bets have been made in turn $t$,
		is less than $3/8$.
		That is $\lambda(\sqs\Theta_i|M_{i+1})\leq 3/8$.
		\item\label{M_capi_small}
		The expected earning for $\bB^{1:n}_{t}$ on the sequences in $M_{i+1}$ that are contained only in
		$(I_{i+1},\phi)$-chubby leaf-restrictions of the Gambler's betting strategies is less than $3$.
	\end{enumerate}

\end{definition}

We will now prove that the Modulo Chooser is a computable winning strategy with residue in the Betting game on open sets against conservative gamblers.
\begin{proof}[Proof of Proposition \ref{key_proposition}]
We begin with the following claim.

	\begin{claim}\label{cl_atomic_increment}
		For any Gambler, at the beginning of the turn $j_i$, when $i$th modulo set is to be chosen, there is a 
		modulo set $M_i=(I_i,m,o_i)$ such that
		the set of positions $I_i$ has properties
		\ref{atomic_large},\ref{lean_small} and the remainder $o_i$ has properties
		\ref{M_lean_small},\ref{M_capi_small}.
%
	\end{claim}
	\begin{proof}
		We prove by induction.
		In the first turn $I_1=[1,\ell], o_1=0$ and the chosen modulo set is $M_1=\Mod(I_1,m,o_1)$.
		Since the Gambler did not define any bets yet, all of the Kolmogorov-Loveland betting strategies are initial, and both
		$\Psi_0$ and $\Theta_0$ are empty.
		We have that
		\ref{lean_small} is satisfied since $\lplus(\Theta_0)=0$
		\ref{atomic_large} is satisfied since $\lplus(\Psi_0)=0$, $z_0=0$
		and $|I_i|=\ell$.
		Properties 
		\ref{M_lean_small} and \ref{M_capi_small} are also satisfied since
		$\lambda(\tilde\Theta_0|M_1)=0$ and $\exc{\bB^{1:n}_0}<1$.
		
		For turns $t\in[j_i,j_{i+1})$, denote with $\Delta_t$
		the multi-set of leaf restrictions of strategies in
		$\bB^{1:n}_t$ that were $(I_i,\phi)$-slimmed down by bets defined in turns
		$j_i,\dots,t$.
		If the $(i+1)$st modulo set is to be chosen in turn $j_{i+1}$, then by definition, 
		the sum-size of $\Delta_{j_{i+1}-1}$  is more than $3/8$.
		For brevity, denote with $\Delta^i$ the multi-set $\Delta_{j_{i+1}-1}$.
		
		Assume that in turn $j_i$ the $i$th chosen modulo set $M_i$ is defined on $I_i$ that has properties
		\ref{atomic_large},\ref{lean_small},
		with a remainder $o_i$ that has properties
		\ref{M_lean_small}, \ref{M_capi_small}.
		
		We next show that there is $I_{i+1}$ with properties \ref{atomic_large},\ref{lean_small}.
		
		In case $\Theta_i \leq 1/4$ then the subset of restrictions in $\Delta^i$ that restrict 
		$I_i$ entirely has sum-size at least $1/8$. 
		The set of positions $I_{i+1}=I_i$ has property \ref{lean_small}
        since $\Theta_i \leq 1/4$, and property \ref{atomic_large} since
        $\lplus(\Psi_i)\geq\lplus(\Psi_{i-1})+1/8$.
		
		In the other case, when $\Theta_i>1/4$, 
		we can use Lemma \ref{smalling_slim_larging_restricted}.
		Let the sum-size of restrictions in $\Delta^i$ that restrict
		$I_i$ entirely be $y$. 
		Let $R$ be the multi-set of leaf restrictions of strategies in $\bB^{1:n}_t$. 
		We have that
		$\lplus(R)=n$.
		Let 
		$I=I_i$,
		$x=\lplus(\Psi_{i-1})+y$, 
		$\delta=1/4$,
		$\delta'=\delta(1-2\delta)=1/8$
		$g=\lceil 8(n-x) \rceil$,
		$q=\delta/\phi=\frac{1}{4\phi}$.
		By assumption, $I_i$ has property \ref{atomic_large} and we have
		$x\geq\lplus(\Psi_{i-1})=z_{i-1}/8\geq(i-1)/8$ and 
		$|I|\geq(\frac{1}{4\phi})^{z_{i-1}}\ell\geq(\frac{1}{4\phi})^{8x}\ell$.
		Since $\ell=(4\phi)^{8n+3}$ we have
		$|I|\geq(4\phi)^{8(n-x)+3}$, and since
		$g\leq 8(n-x)+1$, we have
		$|I|\geq (1/q)^{g+2}$.
		We have that $R,I,\phi,\delta,\delta',x,g,q$
		satisfy the properties of Lemma \ref{smalling_slim_larging_restricted}.
		We also have that there is some positive integer $k\leq g$ and $I'\subseteq I$
		such that
		$|I'|\geq q^k|I|$
		and the sum-size of restrictions in $R$ that are $(I',\phi)$-lean
		is at most $\delta$,
		while the sum-size of restrictions in $R$ that restrict $I'$
		is at least $x+k/8$.
		Let $I_{i+1}=I'$.
		Since $\delta=1/4$, $I_{i+1}$ has property
		\ref{lean_small}. 
		We have $z_i \geq 8x+k\geq z_{i-1}+k$,
		and since $k\geq1$, and $z_{i-1}\geq i-1$, $z_i\geq i$, that is,
		$\lplus(\Psi_i)\geq i/8$.
		Furthermore,
		$|I_{i+1}|
		\geq (\frac{1}{4\phi})^k|I_i|
		\geq(\frac{1}{4\phi})^{k+z_{i-1}}\ell
		\geq (\frac{1}{4\phi})^{z_{i}}\ell$,
		and
		$I_{i+1}$ has property \ref{atomic_large}.
		
		We conclude that in both cases we can find $I_{i+1}$ with properties \ref{atomic_large},\ref{lean_small}.
		
		Let $c=3/2$.
		By Lemma \ref{good_mod} 
		there is some remainder $o_{i+1}$ such that for the modulo set 
		$M_{i+1}=\Mod(I_{i+1},m,o_{i+1})$ we have
		$\exc{\bB^{1:n}_t}(M_{i+1})\leq 3$ and $\lambda(\tilde{\Theta}_i|M_{i+1})\leq 3/8$.
		That is, $M_{i+1}$ has properties \ref{M_lean_small},\ref{M_capi_small}.
	\end{proof}

\begin{claim}\label{cl_finite_choices}
	The Modulo Chooser chooses at most $8n+1$ modulo sets.
\end{claim}
\begin{proof}
	Suppose that the modulo set chooses $8n+1$ modulo sets.
	For the modulo set $M_{8n+1}=\Mod(I_{8n+1},m,o_{8n+1})$,
	by property \ref{atomic_large} the sum-size of the multi-set of leaf restrictions
	that restrict $I_{8n+1}$ entirely is $n$.
	That is, all of the leaf restrictions of strategies in $\bB^{1:n}_{j_{8n+1}-1}$
	restrict $I_{8n+1}$ entirely.
	But then, it is not possible to slim down any of those restrictions, and the condition for the next modulo set to be chosen can never be fulfilled.
\end{proof}

\begin{claim}\label{cl_M_size}
	The measure of every chosen modulo set is $1/8$-approximately $\frac{1}{m}$
\end{claim}
\begin{proof}
		For all $i$, the $i$th modulo set $\Mod(I_i,m,o_i)$ is defined on set of positions $I_i$ that by property \ref{atomic_large} has size of more than
	$(\frac{1}{4\phi})^{z_i}\ell$.
	Since the sum-size of leaf restrictions of $\bB^{1:n}_t$, for any $t$, is at most $n$, $z_i$ is at most $8n$, and we have that 
	$|I_i|\geq (\frac{1}{4\phi})^{8n}\ell=(4\phi)^3\geq\phi=64m^2$.
	Let $\xi=1/8$ so that $|I_i|\geq(\frac{m}{\xi})^2$. 
	We can now use Corollary \ref{corr_M_size} to bound the measure of the chosen modulo sets.
\end{proof}
\begin{claim}\label{cl_size}
	The sum of measures of the sets chosen by the Modulo Chooser with size parameter $k$ 
	is less than $2^{-k}$
\end{claim}
\begin{proof}
	By Claim \ref{cl_finite_choices} the Modulo Chooser chooses at most $8n+1$ modulo sets.
	By Claim \ref{cl_M_size}, the measure of every chosen modulo set is less than
	$\frac{8}{7}\frac{1}{m}$.
	The sum of measures of the chosen sets is then less than $(8n+1)\frac{7}{8}\frac{1}{m}$.
	By definition $n=4+\log m$ and $m=2^{2(k+4)}$, therefore 
	
	$(8n+1)\frac{8}{7}\frac{1}{m}\\
	=\frac{8(8(4+2(k+4))+1)}{7}2^{-2(k+4)}\\
    =\frac{2^7k+2^8+2^9+2^3}{7}2^{-2(k+4)}\\
    <\frac{2^7k+2^{10}}{4}2^{-2(k+4)}
	=[(2^5k+2^8)2^{-k-8}]2^{-k}$.
	This is less than $2^{-k}$, for all $k$.
\end{proof}

\begin{claim}\label{cl_win_last_choice}
	For any conservative Gambler, 
	if the Modulo Chooser chooses finitely many modulo sets in the entire Betting game, 
	the last chosen modulo set contains a subset of size $2^{-n}$ on which none of the first $n$
	betting strategies achieve capital larger than the bound determined by the capital parameter $H$.
\end{claim}
\begin{proof}
	Let $M_i=\Mod(I_i,m,o_i)$ be the last chosen modulo set, chosen in turn $j_i$.
	Since the Modulo Chooser would choose another modulo set if in some turn
	$t\geq j_i$ the sum-size of $\Delta_t$ was larger than $3/8$ we have that
	for all $t\geq j_i$, $\lplus(\Delta_t)<3/8$.
	
	For some $t\geq j_i$, let $M'_i=M_i\setminus(\tilde\Delta_t\cup\tilde\Theta_{i-1})$,
	that is, $M'_i$ is the modulo set $M_i$ without sequences contained in leaf restrictions of strategies in $B^{1:n}_t$ that were slimmed down by bets defined in turns $j_i,\dots,t$ and without sequences in leaf restrictions of strategies in $B^{1:n}_{j_i-1}$ that are $(I_i,\phi)$-lean.
	By Lemma \ref{KL_eta}, since $\phi>m^2$,
	$\exc{\bB^{1:n}_t}(M'_i)
	\leq
	\frac{1}{(1-2\xi)(1-\lambda(\tilde\Theta_{i-1}|M_i))-\lplus(\Delta_t)}
	\exc{\bB^{1:n}_{j_i-1}}(M_i)$, where $\xi=m/\sqrt{\phi}=1/8$.
	We have $\lplus(\Delta_t)<3/8$,
	by property \ref{M_lean_small} $\lambda(\tilde\Theta_{i-1}|M_i)\leq 3/8$
	and by property \ref{M_capi_small} $\exc{\bB^{1:n}_{j_i-1}}(M_i)\leq 3$.
	Therefore,
	$\exc{\bB^{1:n}_t}(M'_i)
	\leq
	\frac{3}{3/4(1-3/8)-3/8}
	=32$.
	Furthermore, by inequalities \eqref{eq_M_M'_measure},\eqref{eq_D|M_D} in Lemma \ref{KL_eta},
	the measure of $M'_i$ is more than 
	$\frac{1}{8}\lambda(M_i)$.
	Then by Claim \ref{cl_M_size},
	$\lambda(M'_i)\geq\frac{1}{8}\frac{7}{8}\frac{1}{m}=\frac{7}{64m}$.
	By Lemma \ref{lo_cap_sub}, 
	$M'_i$ has a subset of size $2^{-n}$ such that for every $j\leq n$ and
	every sequence in the subset, the leaf restriction of the $j$th betting strategy
	at the end of turn $t$ that contains the sequence has capital less than
	$2^{j}\frac{\frac{7}{64m}}{\frac{7}{64m}-2^{-n}}32$.
	As $n=4+\log m$, this is equal to
	$2^{j}\cdot\frac{3}{7}\cdot32$ which is less than $2^{j+4} $.
	Finally, since the Gambler is conservative, this implies that
	$M_i$ contains a subset of size $2^{-n}$ such that
	for any sequence in the subset and
	for every $t$ and every $j\leq n$, the $j$th betting strategy after turn $t$, achieves maximal capital
	smaller than $2^{j+4}+2$ when betting on the sequence.
	This is less than $h_j=2^{j+5}$.	
\end{proof}
	
We have shown in Claim \ref{cl_size} that the Modulo Chooser chooses a set of sequences that is smaller than $2^{-k}$.
By Claim \ref{cl_finite_choices} only finitely many modulo sets are chosen throughout the game. Then by Claim \ref{cl_win_last_choice} and Definition \ref{def_cwswr}, 
the Modulo Chooser has a computable winning strategy with residue in the Betting game on open sets against conservative Gamblers.
\end{proof}

We can now prove the main result.
\begin{proof}[Proof of Theorem \ref{main_result}]
	The main result, Theorem \ref{main_result}, follows from the key proposition, Proposition \ref{key_proposition}, by lemmas \ref{lem_cwswr=>cws}, \ref{lem_cws4conservative=>cws4Basic}, \ref{lem_cws4Basic=>main_res}.
\end{proof}

\bibliography{FBGbibliography}

\end{document}